JIMEI SHEN
YIHAN MO
CHRISTOPHER PLIMPTON
MUSTAFA KAAN BASARAN

# AI in Asset Management and Rebellion Research

*Artificial Intelligence (AI) Investing has taken off over the last 10 years and Rebellion Research is a pioneer in the field. Of course, we have branched out to much more. But, building Machine Learning (ML) is the heart of who we are!*

— Alexander Fleiss, CEO of Rebellion Research (Fleiss, 2021)

On October 30th, 2021, Rebellion Research's CEO announced in a Q3 2021 Letter to Investors that Rebellion's AI Global Equity strategy returned +6.8% gross for the first three quarters of 2021. "It's no surprise", Alex told us, "our Machine Learning global strategy has a history of outperforming the S&P 500 for 14 years" (see Exhibit 1). In 2021, Rebellion's brokerage accounts can be opened in over 70 countries, and Rebellion's research covers over 50 countries. Besides being an AI asset management company, Rebellion also defines itself as a top-tier, global machine learning think tank.

Rebellion Research pays special attention to the financial application of ML and AI and has become a leading figure in the financial technology industry. In addition, through numerous partnerships with top American universities such as Columbia University, New York University,

The authors have no conflict of interest to declare.
The authors acknowledge Professor Alex Tuzhilin for helpful instructions, suggestions, and comments. The authors also want to thank Alexander Fleiss, CEO of Rebellion Research, for providing useful information.

University of California, Berkeley, and Cornell University, Rebellion continues to actively participate in cutting-edge research projects and attract top talent from around the world. A team of dedicated graduate and doctoral students, professors, executives, scientists, and policy experts produced hundreds of reports every year and was often cited by leading scholars and institutions. In addition to research on finance, Rebellion has regularly worked with U.S. government and military experts in recent years, covering areas ranging from drone image recognition and retail inventory analysis. "Our in-house expertise, cutting-edge artificial intelligence analysis platforms, and close collaboration with the public and private sectors combine to generate influential insights into global events through in-depth research, data science and analysis, and the application of data-driven modeling", said Alex.

Alex planned to build a Rebellion ML & AI ecosystem. Should Rebellion stay in the asset management area or jump into other areas? How could the Rebellion strategically move towards a more broad area? What were Rebellion's new or alternative business models?

## Rebellion Research History

Rebellion Research was founded in 2003 by Alexander Fleiss, Jonathan Sturges, Jeremy Newton, and Spencer Greenber (Patterson, 2010). Alexander, Jonathan, and Jeremy met at a math class at Amherst College, and Spencer from Columbia University joined afterward (Farzad, 2006). Rebellion was one of the first Wall Street firms to use AI (Young, 2016) and the first to launch the AI-only fund (Guangfa Securities, 2017).

Initially, Rebellion used Bayesian Networks to make market predictions (Comstock, 2010). Rebellion's AI system is based on the Bayesian algorithm, which analyzes the macro, industry, and company levels. The model can automatically integrate historical data with the latest data to automatically predict market trends. The first AI investment fund launched by the company in 2007, based on Bayesian machine learning, successfully predicted the 2008 stock market crash and gave a rating of F on Greek bonds in September 2009. At that time, Fitch's rating was still A, and Rebellion downgraded Greek bonds one month ahead of the official.

In recent years, Rebellion has also begun to gradually apply more technologies to asset management, such as Natural Language Processing (NLP) based market sentiment analysis, 13F data and ML (also known as fundamental quantitative analysis), bitcoin and ML, Reinforcement Learning (RL) for Portfolio Allocation, Machine Learning Portfolio on 13F. "However, ML is our heart", Alex said, "In addition, RL is more useful than Deep Learning (DL) for us".

## AI Technologies in Asset Management

*Background*

Turing (1950) proposed the well-known Turing Test, which set off the first wave of artificial intelligence (Turing and Haugeland, 1950). In 1956, at the summer academic seminar held at Dartmouth University, McCarthy, Minsky, and Shanno formally gave the term Artificial Intelligence (AI) (McCarthy et al., 2006). However, due to the limitations of computer computing power and symbolism theory, the popularity of artificial intelligence faded rapidly.

In the 1980s, artificial intelligence rose again, and the breakthrough originated from abandoning symbolism and using statistical thinking instead. In 1991, IBM Deep Blue defeated chess player Kasparov, which set off the second wave of artificial intelligence. However, artificial intelligence at that time was limited by data volume and test environment and did not have much practical value.

The third wave of artificial intelligence originated from the concept of deep learning put forward by Hinton et al. in 2006 (Hinton, Osindero, & Teh, 2006). At this time, the development of the Internet industry has formed massive amounts of data, and the innovation of the graphics processing unit (GPU) has dramatically improved the computing power of computers. With the maturity of these two conditions, artificial intelligence has achieved rapid development. Taking Google AlphaGO's victory over Go player Lee Sedol in 2016 as a landmark event, the concept of AI has been deeply rooted in people's hearts, the theory has become practice, and scientific research achievement has gone to the market.

Similar to the three waves of artificial intelligence, the application of information technology in the financial field could be divided into three stages. Since the 1950s, technology has gradually entered the e-finance period. The development of information systems had promoted the electronization and automation of financial services, improved the service capacity of outlets, and reduced the transaction costs of deposits, loans and remittances. The main commercial applications were credit cards, ATMs, POS machines, CRM systems, etc. In the 1990s, with the generation of massive user data, the scope and depth of financial services were greatly extended. In addition, the enhancement in cyber security transformed financial service mode, resulting in online banking and mobile payment. With the emergence of new technologies since 2016, mainly AI, Blockchain, Cloud Computing, Big Data, the intelligence and personalization of the financial industry was further promoted. Information asymmetry and transaction costs were reduced significantly. The most affected area was likely asset management and fundamental analysis was considered to be the cornerstone of asset management (Bartram, Branke, & Motahari, 2020).

*Machine Learning in Asset Management*

Mittermayer (2004) proposed NewsCATS, a system that classified news and did trading based on the support vector machine (SVM). Choudhury et al. (2008) used text information for communication on Engadget to predict stock price. Results showed that the model could predict the stock price trend of blog-related stocks in the future effectively (the accuracy rate is 78% to 87%). Schumaker and Chen (2009) proposed an SVM-based method for financial news article analytics. It was found that the prediction results of the model including both the terms of the article and the stock price at the publication time were close to the real future stock price (with an MSE of 0.043), the price movement trend was the same as the real future price (with an accuracy of 57.1%), and the maximum return (2.06%). The Arizona Financial Text System (AZFinText), based on machine learning, was created by using a comprehensive method of language, finance, and statistical technology to deal with the problem of discrete prediction of stock prices. The trading return rate of this system was 8.50%, which was 2% higher than the best performing quantitative fund (Schumaker & Chen, 2009a). Li (2010) used Naive Bayes to verify the relationship between the information in the company's public document and the company's stock price, and the results showed the average tone of forward-looking statements had a positive correlation with future earnings. Groth and Muntermann (2011) used news reports and stock price data to determine that the disclosures provided in text data led to the most extraordinary risk exposures. Compared to the nearest neighbor (KNN), neural network (NNnet), naive bayes, the results showed that SVM had the highest AUC (0.642) and accuracy (78.96%). Hagenau et al. (2012) used market feedback as a part of word selection and constructed an SVM model with an accuracy of 65.1% for automatically predicting stock prices using unstructured text information. Li et al. (2014) used the eMAQT model based on the SVM to confirm the relationship between investor behavior and news information. Bogle and Potter (2015) built a stock price prediction model SentAMaL based on ML and emotion analysis and studied the mixed prediction model of emotion expressed on Twitter and machine learning to predict the US index. The decision tree model reached 100% accuracy, and the performance of the decision tree model in this task was better than the SVM and artificial neural network. Piñeiro‑Chousa (2017) uses the Logit model to analyze the performance of investors on social media and its influence of text information on the market. The results showed that the combination of different factors did have an impact on the market, which depended on the type of investors.

*Deep Learning in Asset Management*

Bollen et al. (2011) conducted the Granger causality test and the self-organizing fuzzy neural network, and found that specific public emotion could significantly improve the prediction accuracy of the Dow Jones Industrial Average. Li et al. (2016) proposed a tensor-based TeSIA algorithm, which transformed the stock forecasting model into a high-order tensor regression learning problem. Cabanski et al. (2017) compared the performance of the SVM and recurrent neural network (RNN) by extracting word vectors and emotional dictionary features of financial data. The results showed that the RNN achieved excellent results of 0.729 and 0.702 in SemEval-2017 respectively. Heston and Sinha (2017) used neural networks to measure emotions and found that daily news could only predict stock returns for days of one or two. And positive news would increase stock returns rapidly, while negative news had a long-term reflection period. Most of the response delayed to news occurred near the announcement of subsequent earnings. Kar et al. (2017) used support vector regression and long short-term memory network (LSTM) to establish the application task of target-dependent fine-grained text emotion prediction in the financial field. Ghosal et al. (2017) proposed a hybrid model based on convolution neural network (CNN), LSTM, vector averaging, and feature-driven multilayer perceptron to improve the accuracy of target-dependent text emotion analysis tasks in the field of stock investment. Zhang et al. (2017) proposed a predictable method for the rise and fall of stocks. In the aspect of feature extraction, besides stock price information, it also extracted emotional features related to stocks based on news extraction. It was found that compared with the SVM classifier based on price features, the proposed method has an improvement of more than 5% in the stock rise and fall prediction. Lee and Soo (2018) proposed a recursive convolution neural network that combines the advantages of convolution, sequence model, and knowledge extraction, and can capture information from the trained convolution filter by embedding pre-trained words in the corpus. The results showed that the recurrent convolutional neural network (RCN) model combined with technical analysis indicators predicts that four of the five targets have lower RMSE (TSMC: 28%, Grape Gold: 38%, CSC: 30%, Taiwan Stock Exchange: down 53%). Tan et al. (2019) proposed the application of the tensor-based eLSTM model in financial news, using tensor instead of cascade vector to model market information, and using the event-driven mechanism to balance the heterogeneity of different data types in LSTM. Tan et al. (2019) conducted experiments on the annual data of China's securities market, which showed that the eLSTM method was superior to existing algorithms such as AZfinText, eMAQT, and TeSIA. Araci (2019) introduced the state-of-the-art BERT into the finance area, and the proposed FinBERT outperformed ML models.

*Reinforcement Learning in Asset Management*

   Moody et al. (1998) put the recursive reinforcement learning algorithm model (RRL) and in the tests of the S&P 500 Index and US stocks, the reinforcement learning model with Sterling ratio as the objective function has the highest return. Moody and Saffell (1999) applied RRL to the fields of single stock and asset portfolio. They tested the intraday foreign exchange market (USD/GBP), S&P 500, short-term US treasury bonds, and other financial assets. With the yield as input and the differential Sharpe ratio as the objective function, the experiment was carried out under the condition of transaction cost of 5‰. The return of RRL strategy was higher than Q-learning strategy and buy-hold strategy, and the transaction number was lower than Q-learning strategy. Moody et al. (2001) added the short-position wait-and-see action Ft $\in$ {-1, 0, 1} on the basis of RRL, and Ft=0 means suspending trading and reducing risks within a certain period of time. In addition, the downside deviation ratio was used instead of the Sharpe ratio as the objective function to test the return status of the model when the market goes down. This was the first time that RRL was applied to foreign exchange high-frequency trading between sterling and US dollar. The comparison between RRL and Q learning shows that RRL was superior to Q learning strategy in many aspects, which also proves that RRL was more suitable for high-frequency trading. Gold (2003) proposed to replace the single-layer neural network with the multi-layer neural network in the RRL model. Gold tested it in 25 different high-frequency foreign exchange markets. The test results showed that both single-layer RRL and multi-layer RRL could achieve profitability, and the performance of multi-layer RRL was worse than that of the single layer. Gorse (2011) did similar experiments, attempting to use the multi-layer instead of the single-layer. Experimental results showed that compared with single-layer RRL, the performance of multi-layer RRL was not significantly improved. Jiang et al. (2017) established a deep reinforcement learning (DRL) portfolio management framework. Despite a 0.25% commission rate, the framework proved able to gain at least 4-fold returns in 50 days in the financial market and 10-fold in cryptocurrency. Benhamou et al. (2020) built an Augmented Asset Management with DRL (AAMDRL), which can achieve superior returns (22.45% in 3 years and 16.42% in 5 years) at a lower risk.

*Summary of AI Technologies in Asset Management*

In this session, we discussed the frontiers of the AI technologies (ML, DL, RL) that Rebellion Research mainly applied. We can find that the usage of AI in asset management, such as portfolio management, trading, portfolio risk management, is an emerging field that attracts both academia and industry. As Alex said, AI technology can effectively discover information that humans cannot or are difficult to capture, process high-latitude information that is difficult for humans to process, and analyze unstructured data (such as news, social platform data, pictures, etc.) to make scientific and rational decision-making.

In addition, we find that the application of AI technology in asset management was closely related to the development of AI technology. The application of AI in the financial field has existed since this concept was put forward, but it was not widely used due to objective factors such as hardware. Since Hinton et al. proposed the deep neural network (DNN) in 2006, the application of AI in the financial field has also changed from shallow learning (SVM, Logit, RRL, etc.) to deep learning (LSTM, BERT, DRL, etc.).

## Competition in AI Asset Management

Established in 2003, Rebellion Research's asset under management (AUM) raised from $2 million when it released the first AI-based fund in 2006 to $21 million in 2021. With the first mover's success, especially Rebellion's success in the 2008 financial crisis, early adopters gradually entered the market around 2008.

*High-Flyer Quant*. In 2008-2014, founders established an AI team to explore fully automated transactions in China. In 2015, High-Flyer Quant was founded, and smart investment was deeply cultivated. Zhejiang Jiuzhang Asset Management Co., Ltd. was established with a private equity fund license. In 2016, the AI algorithm broke through and became a member of the China Fund Industry Association. Established High-Flyer Quantitative Investment Management Limited Partnership with private equity fund license. In 2017, the deep learning model was fully applied by High-Flyer. In 2018, AI technology led the iterative strategy development, explored the integration of multiple strategies, and won the honor of the China Private Equity Golden Bull Award for the first time. In 2019, established a supercomputing center to provide scientific research-level basic computing power, accelerated the research of complex neural networks, and established High-Flyer Capital Management (Hong Kong) Limited. High-Flyer had about $4.7 billion AUM in 2021.

*D. E. Shaw & Co.* D.E. Shaw, a hedge fund, was founded in 1988 in New York. In 2018, D.E. Shaw announced the foundation of a new machine learning group that designs, implements, tests, and scales up new models for asset pricing, trading signal development, and other financial applications. D.E. Shaw's AUM raised significantly from 2018's $17 billion to more than $60 billion in 2021.

*Man Group*. Man Group, an active management business company, was founded in 1783 in London that reported $117.7 billion AUM. In 2017, the Man Group announced that they began letting computers trade with $43 billion assets through quantitative trading.

*Bridgewater Associates*. Bridgewater Associates, an investment management firm, was founded in 1975 in New York. In 2012, David Ferrucci, who led IBM Waston's development of the supercomputer that beat humans at Jeopardy in 2011, joined Bridgewater as AI lead. Bridgewater reported $140 billion AUM in March 2021.

*Citadel LLC*. Citadel, a hedge fund, was founded in 1990 in Chicago. In 2017, Citadel hired a new head of AI, Li Deng, previously chief scientist of AI and partner research manager at Microsoft. On March 31, 2021, Citadel reported a $33.1 billion AUM.

*Two Sigma Investments*. Two Sigma Investments, a hedge fund, was founded in 2001 in New York. In 2016, Two Sigma introduced an AI challenge named Halite that used coding to build smart trading bots. In 2018, Two Sigma hired Google Brain's scientist Mike Schuster to expand its AI business as a $52 billion quant hedge fund. Two Sigma had an AUM of more than $60 billion in 2021.

*Point72 Asset Management*. In 2014, the S.A.C. Capital, a hedge fund, was founded in 1992 in New York converted to Point72. On Nov 15, 2021, Point72 raised $600 million for an AI-assisted private equity fund. Point72 had $23.1 billion AUM in 2021.

## Can Do vs Should Do

In the current AI Hype, people are trying to apply AI-based solutions to various fields, and the asset management industry is also actively exploring the use of AI systems, and taking advantage of its data insight capabilities to provide better services.

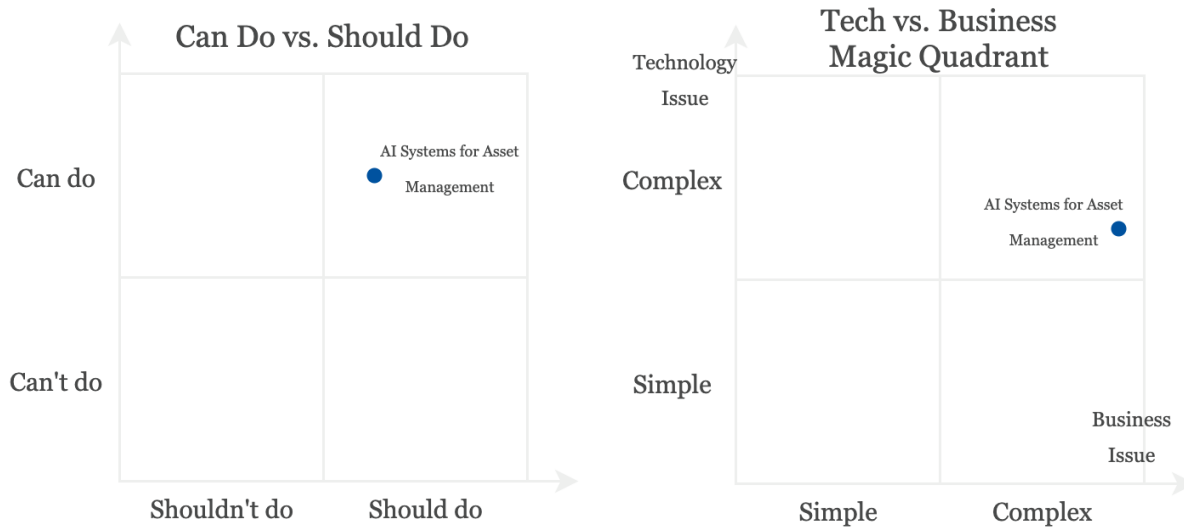

We believe that in the field of asset management, the current AI systems do provide data insights far beyond human beings' ability. They can better capture the detailed information in the data and automatically discover and learn the laws of market changes caused by this information. Therefore, we believe that technologically, though the AI systems are costly and complex, the performance of some pioneers in this area has proven the value of these systems. The use of AI systems to optimize asset allocation is a "can do" decision.

But on the other hand, the risk of AI systems in the asset management field has not yet been fully tested, and customers' trust in a "black box" system that is difficult to explain the logic of decision has not yet been fully established. Therefore, we believe that the application of AI and asset management is a "should do" decision, but we are more vigilant about its systemic risks.

**Rebellion's Vision**

Rebellion Research was the first mover in AI in asset management. "We are more than AI in asset management - providing AI financial advisors and hedge funds", Alex pictured, "we are a global machine learning think tank".

*AI Investing*. Rebellion Research's technology enabled it to process an extremely diverse set of information, and its analysis was based on macroeconomic, fundamental, technical, and traditional factors, such as growth, value, momentum, and so on. Rebellion used its knowledge of volatility and interrelationships among stocks to create portfolios that balance risk and expected returns. Bayesian statistics were the backbone of Rebellion's artificial intelligence-based investment software, providing a flexible framework that enabled Rebellion to automatically combine new data available every day with previous market knowledge to predict stock performance. Rebellion was founded by three mathematicians who believe they can find alpha by using machine learning for long-term investment, covering 53 countries to individual stock prices, to create an optimized portfolio that beats the global stock market with 24/7 service.

*AI Stock Advising*. Rebellion's AI-based global equity strategy has been managing assets for clients since January 1, 2007. Since its establishment in 2007, it has performed significantly better than the S&P 500 (see Exhibit 1). This strategy held a diversified portfolio consisting of 90-120 global stocks. Rebellion's absolute return strategy was a 0 beta, low volatility strategy consisting of 15-20 ETFs (see Exhibit 2). Rebellion provided cash alternatives for customers, seeking no currency exposure or risk, wanting to add security anchors to their accounts, and seeking low or negative market correlation.

*AI Financial Planning*. Rebellion Research's AI investment plan actively paid attention to the front page news. Rebellion's AI could create and monitor financial databases, categorize financial data, improve financial accounting processes, perform complex analysis, help financial planning, assist in project management by utilizing NLP to analyze 10ks and financial reports. Additionally, Rebellion examined the backbone of the business and how companies survived after the pandemic.

*Rebellion TV*. Rebellion TV brought together the most brilliant brains in ML and AI, providing cutting-edge news in AI. Guest speakers vary from Turing Award winner and Facebook's Chief AI Scientist Professor Yann LeCun, Nobel Prize winners Stanford Professor Paul Romer & MIT Professor Frank Wilczek, Harvard Law School Professor Alan Dershowitz, Wharton Professor and Microsoft Azure Chief Economist Amit Gandhi, etc. to ex Compliance Chief of JOM Eric Young, Tractable AI CEO on Computer Vision Alex Dalyac, Mckinsey's Head of AI Ernst & Young's Head of AI & Founder Pascal Bornet, Head of Machine Learning Strategies of JP Morgan Peng Cheng, etc.

*Research*. Rebellion's research covered a broad range, including auto, aviation and transportation, AI & ML, cryptocurrency & blockchain, cyber security and hacking, education, military, sustainable investing, space, trading and investing, technology. Their research team consisted of a group of undergraduate, graduate, and Ph.D. students from world-first universities such as NYU, MIT, UCB, Amherst College, etc.

*Education*. Rebellion also had close relationships with many high-profiled universities in AI in asset management education, such as UCB Haas MFE, MIT Sloan MFin, Cornell SC Johnson, NYU Courant Finmath, etc., in providing high-quality guest talks and co-op internships.

## Challenges & Risks

*Technologies*. Although AI had been very powerful in asset management and other finance areas, it had its limitations. DL-based AI systems have the ability to learn massive amounts of data that humans cannot process and capture detailed information that traditional methods cannot use, but based on the current technology, deep learning models are still a "black box" for users. As human financial advisors, when we constructed investment portfolios, we could explain the reasoning and logic behind them. However, robo advisors, DL-based especially, were poor at interpretation, the evaluation of the model's effect strongly relies on historical data, and we cannot predict whether the model's output is still valid under unconventional input, which reduced the liability, These feedbacks may be coupled with other AI models in the market and cause further market volatility. Additionally, despite the fast development of hardware, it was still hard to process hundreds of tons of finance news and transaction data in real-time. This response time of the system could be acceptable when the market is relatively stable, but the reliability of AI-based systems in a crash or crisis is still waiting to be valid.

*Threats from Fast-movers*. Rebellion was faced with many competitors, not only new AI asset management companies, but also previous traditional hedge funds which were transforming to AI. The former was usually featured and founded by previous famous fund managers or university researchers. The latter, with hundreds of billions of AUM and AI experts from IBM Waston, Microsoft Research, Google Brain, was generally leading and taking over the AI asset management market.

## Rebellion's Next Move

In light of Rebellion's previous success in the 2008 financial crisis, 2009 greek debt crisis, 2014 oil price crash, 2019 COVID-19, etc., Alex was quite confident in Rebellion Research's AI system. Alex's ambition was far more than the finance area, "we also routinely collaborate with US government and military experts, covering domains such as drone image recognition and retail inventory analysis."

On the technology side, ML, more accurately, Bayes, was the heart of Rebellion. With the new development, breakthroughs, and trends in DL, RL, and Auto-ML, should rebellion reconstruct the existing system with more deep methods?

On the business side, should Rebellion keep its main focus on the finance area or explore more broad areas in AI? How could rebellion survive the fast-movers?

**Exhibit 1** Performance Chart of Rebellion's Global Equity Strategy

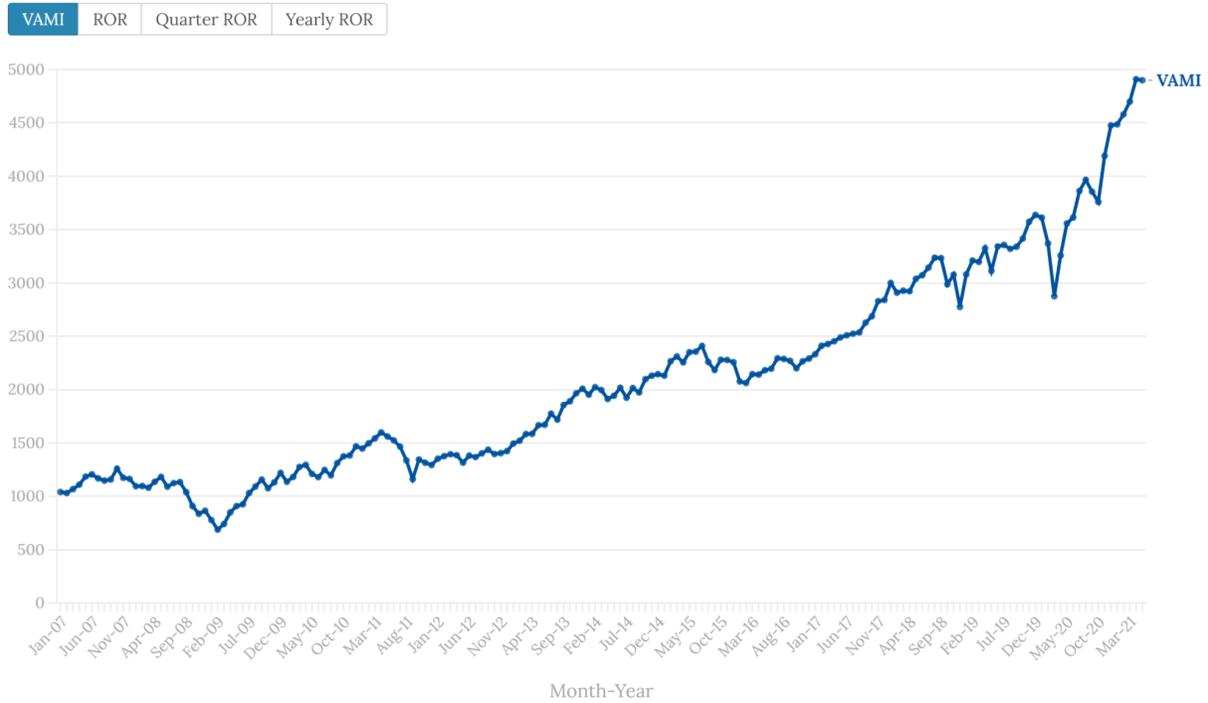

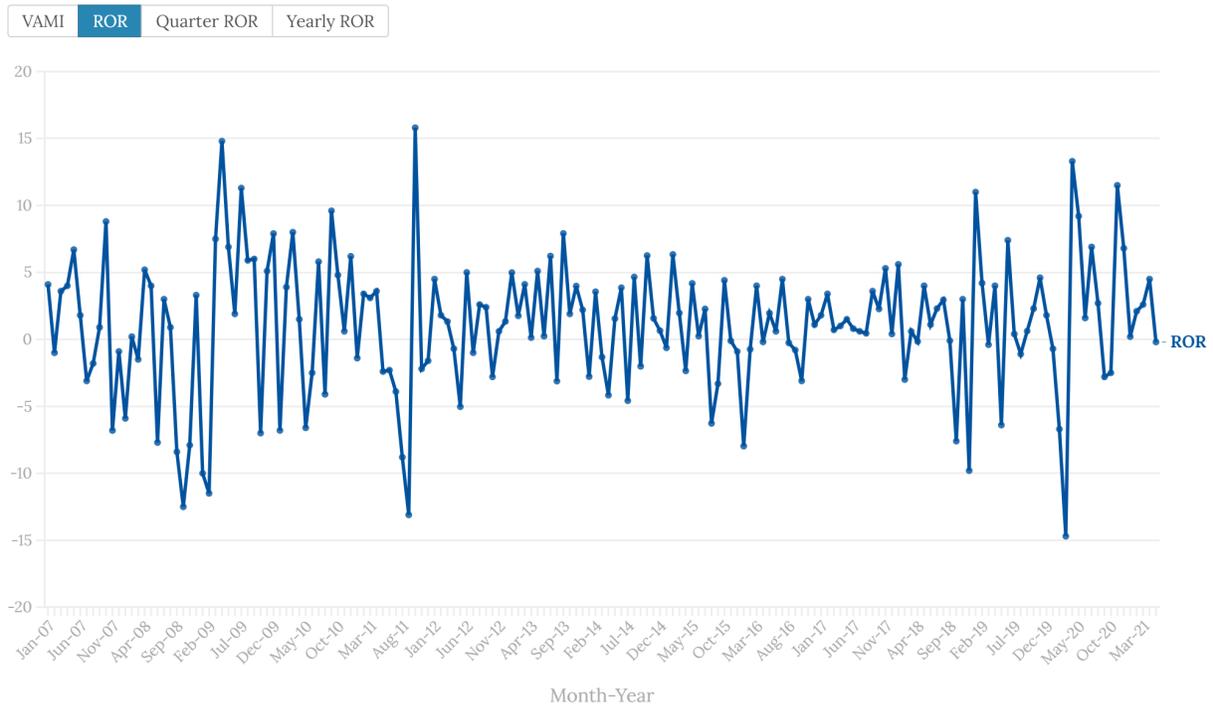

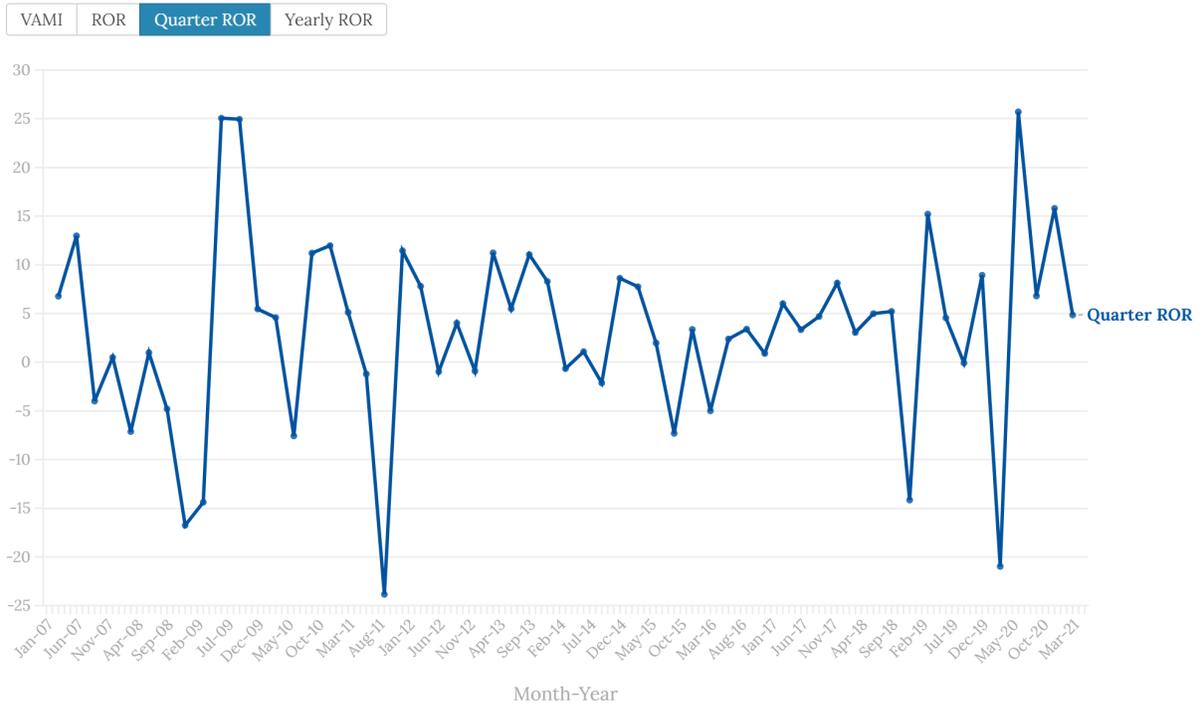

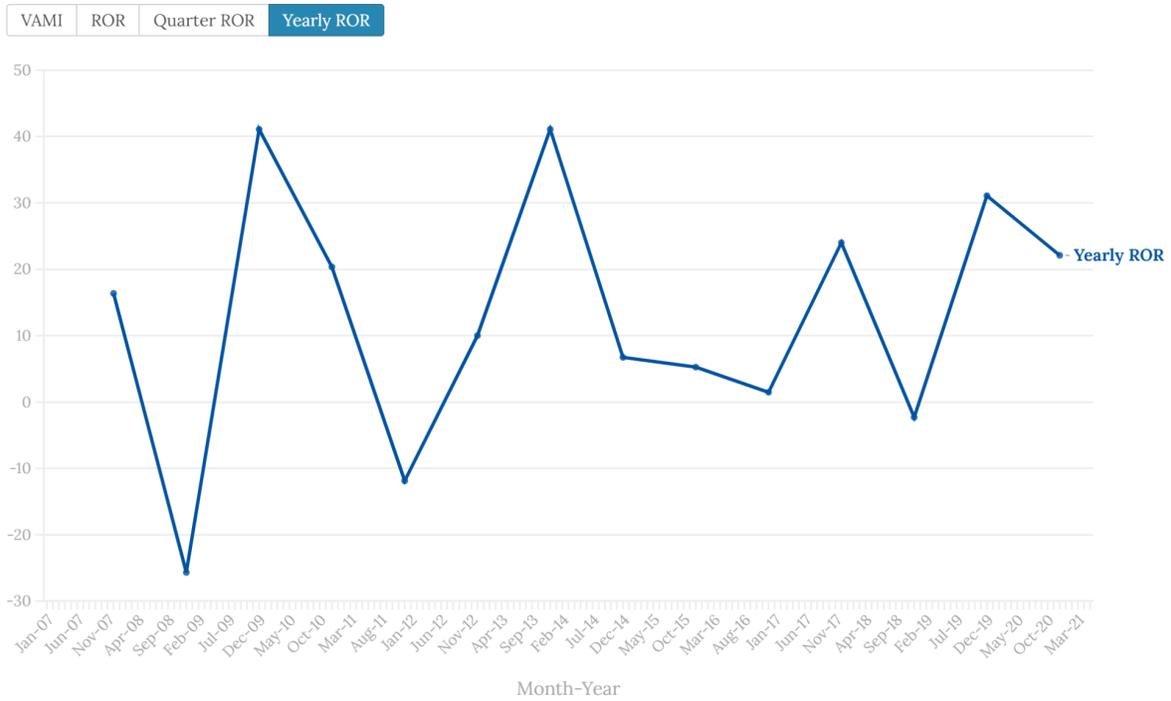

Source: Rebellion Research. https://www.rebellionresearch.com/

**Exhibit 2** Performance Chart of Rebellion's Absolute Return Strategy

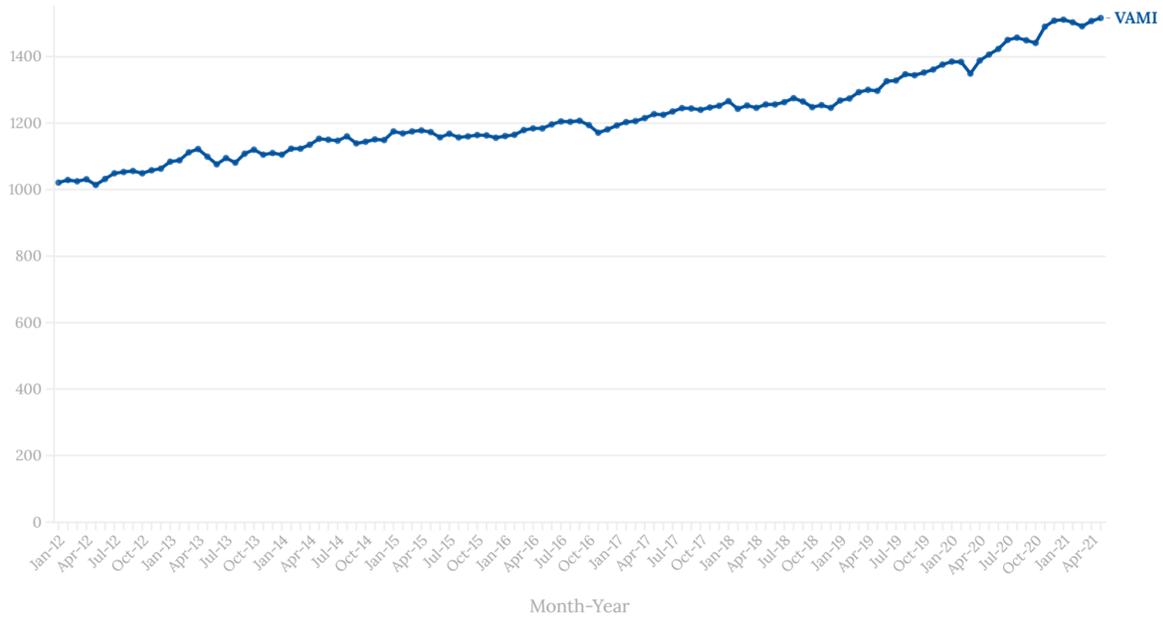

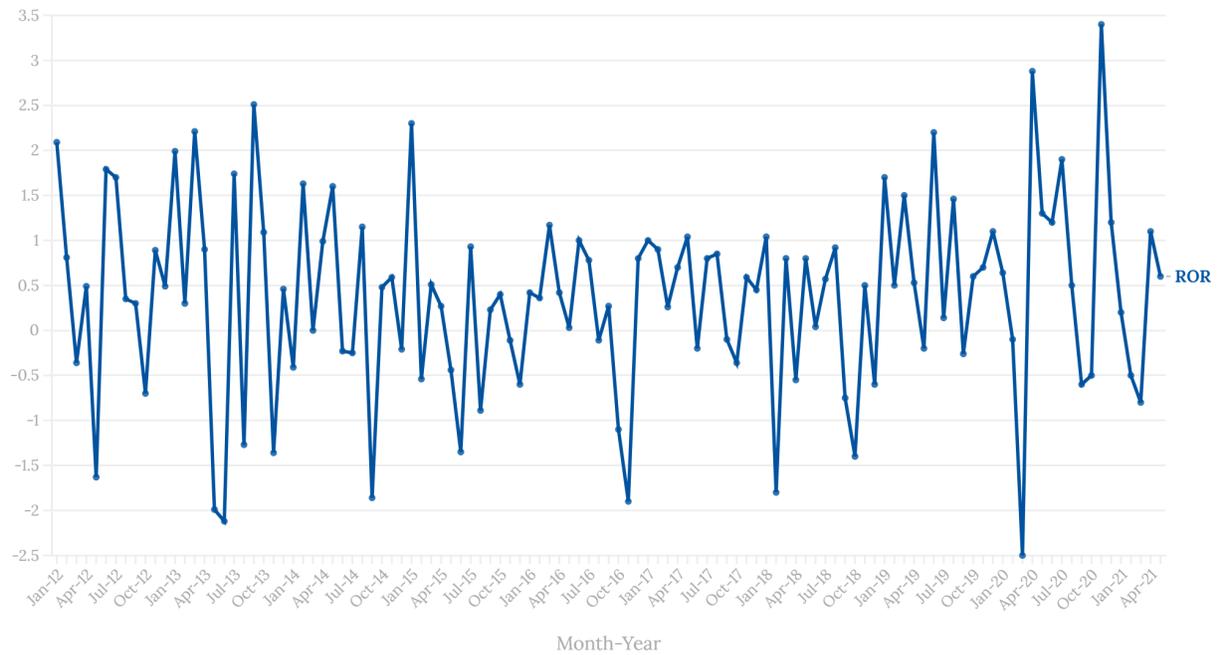

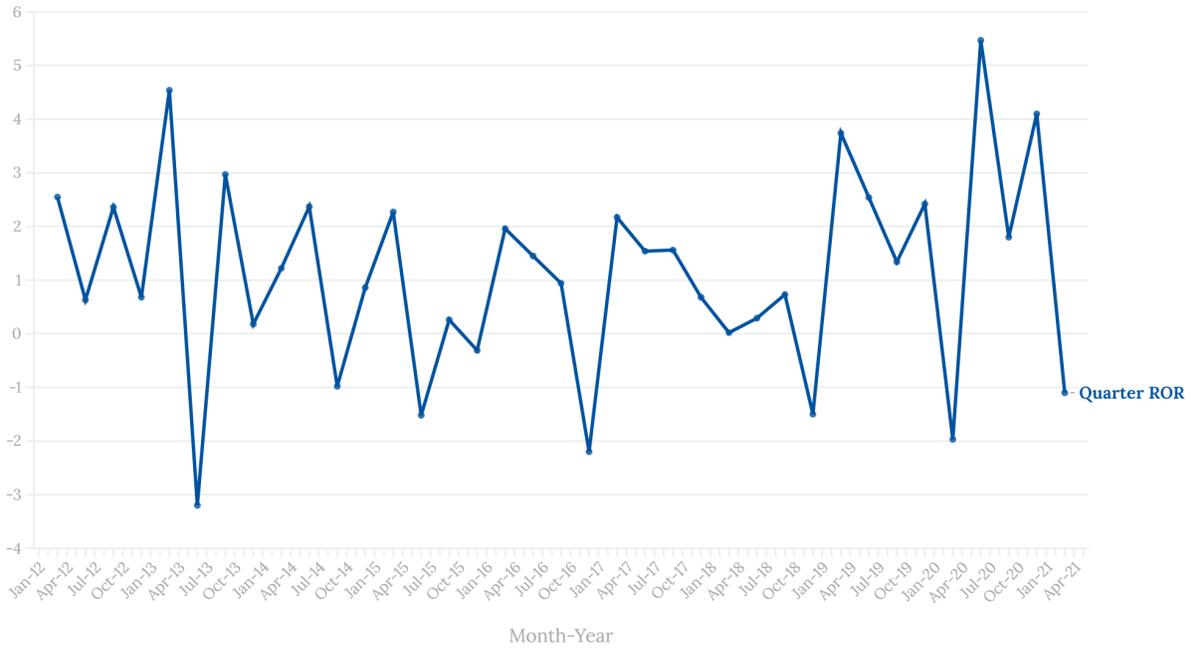

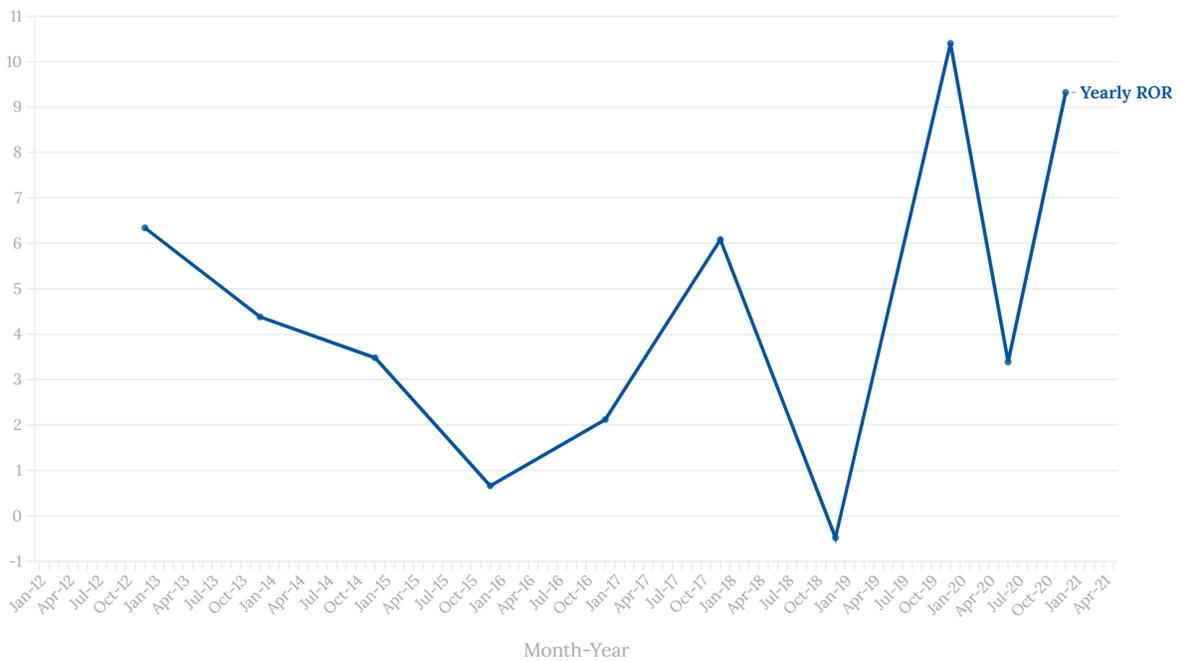

Source: Rebellion Research. https://www.rebellionresearch.com/

**Endnotes**